\documentclass[prc,twocolumn,nofootinbib,floatfix,showpacs]{revtex4}
\usepackage{graphics}
\usepackage{epsfig}
\usepackage{amsfonts}
\usepackage{amsmath}
\usepackage{bm}% bold math

\usepackage{color}
\def\bra#1{\langle #1\vert}
\def\ket#1{\vert #1\rangle}
\def\ad#1{a^+_{#1}}
\def\ovlap#1#2{\langle #1\vert #2 \rangle}

\begin{document}
\title{Giant Monopole Resonances and nuclear incompressibilities
studied for the zero-range and separable pairing interactions.}
\author{P.~Vesel\'{y},$^1$ J.~Toivanen,$^1$  B.~G.~Carlsson,$^2$ J.~Dobaczewski,$^{1,3}$
N.~Michel,$^1$ and A.~Pastore$^4$}
\affiliation{$^1$Department of Physics, University of Jyv\"{a}skyl\"{a},
P.O. Box 35 (YFL) FI-40014, Finland}
\email{petr.p.vesely@jyu.fi}
\affiliation{$^2$Division of Mathematical Physics, LTH, Lund University,
P. O. Box 118, S-22100 Lund, Sweden}
\affiliation{$^3$Institute of Theoretical Physics, Faculty of Physics,
University of Warsaw, ul. Ho\.{z}a 69, PL-00-681 Warsaw, Poland}
\affiliation{$^4$Universit\'e de Lyon, F-69003 Lyon, France; Universit\'e Lyon 1,
43 Bd. du 11 Novembre 1918, F-69622 Villeurbanne cedex, France\\
CNRS-IN2P3, UMR 5822, Institut de Physique Nucl{\'e}aire de Lyon}

\date{\today}

\begin{abstract}
\begin{description}
\item[Background:] Following the 2007 precise measurements of monopole
strengths in tin isotopes, there has been a continuous theoretical
effort to obtain a precise description of the experimental
results. Up to now, there is no satisfactory explanation of why the
tin nuclei appear to be significantly softer than $^{208}$Pb.

\item[Purpose:] We determine the influence of finite-range and
separable pairing interactions on monopole strength functions in
semi-magic nuclei.

\item[Methods:] We employ self-consistently the Quasiparticle Random
Phase Approximation on top of spherical Hartree-Fock-Bogolyubov
solutions. We use the Arnoldi method to solve the linear-response
problem with pairing.

\item[Results:] We found that the difference between centroids of
Giant Monopole Resonances measured in lead and tin (about 1\,MeV)
always turns out to be overestimated by about 100\%. We also found
that the volume incompressibility, obtained by adjusting the
liquid-drop expression to microscopic results, is significantly
larger than the infinite-matter incompressibility.

\item[Conclusions:] The zero-range and separable pairing forces
cannot induce modifications of monopole strength functions in tin to
match experimental data.

\end{description}
\end{abstract}


\maketitle

\section{Introduction}
The incompressibility of infinite nuclear matter as well as of finite
nuclei has been studied in a number of theoretical papers and
reviews. In the classic review by Blaizot~\cite{[Bla80]} the
connection between the finite-nucleus incompressibility and
centroid of the Giant Monopole Resonance (GMR) was shown. This
relation allows us to study incompressibility of nuclei through
microscopic calculations of the monopole excitation spectra.  It
also brings us the possibility to directly compare theoretical
results with experimental data.  For examples, see the measurements
presented in Refs.~\cite{[You99],[Li07],[Li10]}.

In Ref.~\cite{[Pie10]}, it was shown that the self-consistent models
that succeed in reproducing the GMR energy in the doubly-magic
nucleus $^{208}$Pb systematically overestimate the GMR energies in
the tin isotopes. In spite of many studies related to the
isospin~\cite{[Sag07],[Pea10],[Cen10]}, surface~\cite{[Sha09]}, and
pairing~\cite{[Civ91],[Li08a],[Nik08a],[Tse09],[Kha09a],[Kha09b],[Kha10]}
influence on the nuclear incompressibility, to date there is no
theoretical explanation of the question "Why is tin so
soft?"~\cite{[Pie07],[Pie10]}. For an excellent recent review of the
subject matter we refer the reader to Ref.~\cite{[Li10]}.

Studies in Refs.~\cite{[Kha09a],[Kha09b]} were restricted to the
effect of zero-range pairing interaction. In the present paper we
focus on a different kind of  pairing force, namely, we
implement the finite-range, fully separable, translationally
invariant pairing interaction of the Gaussian form
\cite{[Dug04],[Tia09],[Nik10]}, together with the general
phenomenological quasilocal energy density functional in the
ph-channel \cite{[Car08e]}.  We have performed calculations for all
particle-bound semi-magic nuclei starting from $Z=8$ or $N=8$, up to
$Z=82$ or $N=126$. The ground-state properties were explored within
the Hartree-Fock-Bogolyubov (HFB) method, whereas the monopole
excitations were calculated by using the Quasiparticle Random Phase
Approximation (QRPA) within the Arnoldi iteration
scheme~\cite{[Toi10]}. For the numerical solutions, we used an
extended version of the code HOSPHE~\cite{[Car10d]}.

The paper is organized as follows. In Secs.~\ref{sec2}
and~\ref{separable}, we briefly outline the Arnoldi method to solve
the QRPA equations and present the separable pairing interaction,
respectively. In Sec.~\ref{sec4}, we discuss the nuclear
incompressibility, including its theoretical description, definitions
in finite and infinite nuclear matter, and relations to monopole
resonances. Then, our results are shown and discussed in
Sec.~\ref{sec5} and conclusions are given in Sec.~\ref{sec6},
whereas the Appendix presents numerical tests of the approach.

\section{QRPA method}
\label{sec2}

In the present study, we solve the QRPA equations by using the
iterative Arnoldi method, implemented in Ref.~\cite{[Toi10]}. It
provides us with an extremely efficient and fast way to solve the
QRPA equations. The QRPA equations are well known
\cite{[RS80],[Bla86]} and have been recently reviewed in the context
of the finite amplitude method~\cite{[Avo11]}. Therefore, here we
only give a brief resum\'e of basic equations, by presenting their
particularly useful and compact form.

Basic dynamical variables of the QRPA method are given
by the generalized density matrix ${\cal R}$,
\begin{equation}
 {\cal R} = \begin{pmatrix} \rho & \kappa  \\ \kappa^{+} & 1-\rho^{T} \end{pmatrix}
 = \begin{pmatrix} V^*V^T & V^*U^T \\ U^*V^T & U^*U^T \end{pmatrix} ,
 \label{R-dens}
\end{equation}
corresponding to mean-field Hamiltonian
${\cal H}= \partial {\cal E}/ \partial {\cal R}$,
 \begin{equation}
 {\cal H} = \begin{pmatrix} h-\lambda & \Delta  \\ \Delta^{+} & -h^{*}+\lambda \end{pmatrix} .
 \label{H-dens}
\end{equation}
The standard HFB equations that define amplitudes $U$ and $V$ read
\begin{equation}
  \begin{pmatrix} h-\lambda & \Delta  \\ \Delta^{+} & -h^{*}+\lambda \end{pmatrix} \begin{pmatrix} U & V^*  \\ V & U^* \end{pmatrix} = \begin{pmatrix} U & V^*  \\ V & U^* \end{pmatrix} \begin{pmatrix} E & 0  \\ 0 & -E \end{pmatrix}
 \label{HFB-eq}
\end{equation}
where the diagonal matrix $E$ contains positive quasiparticle energies.
Then the quasiparticle ($\chi$) and quasihole ($\varphi$) states
are given by columns of eigen-vectors:
\begin{equation}
 \varphi := \begin{pmatrix} V^*  \\ U^* \end{pmatrix},  \hspace{2.0cm} \chi := \begin{pmatrix} U  \\ V \end{pmatrix},
\end{equation}
that is,
\begin{equation}
 {\cal H} \varphi = - \varphi E,  \hspace{2.0cm} {\cal H} \chi = \chi E .
\end{equation}

The vibrational time-dependent HFB state $|\Psi(t)\rangle$,
\begin{equation}
 |\Psi(t)\rangle = |\Psi\rangle + \mbox{e}^{i\omega t} |\tilde{\Psi}\rangle ,
 \label{TDHFB}
\end{equation}
where $|\tilde{\Psi}\rangle$ is a small-amplitude correction, leads to
the time-dependent density matrix,
\begin{equation}
 {\cal R}(t) = {\cal R} + \mbox{e}^{i\omega t} \tilde{\cal R} + \mbox{e}^{-i\omega t} \tilde{\cal R}^{+}
\end{equation}
and time-dependent
mean field ${\cal H}(t)$,
\begin{equation}
 {\cal H}(t) = {\cal H} + \mbox{e}^{i\omega t} \tilde{\cal H} + \mbox{e}^{-i\omega t} \tilde{\cal H}^{+} .
\end{equation}
After a linearization of fields in the time-dependent Hamiltonian,
one obtains the QRPA equations in a simple form,
\begin{equation}
 -\hbar \omega \tilde{\cal R} = [{\cal H},\tilde{\cal R}] + [\tilde{\cal H},{\cal R}].
 \label{QRPA1}
\end{equation}

In this approach, states in Eq.~(\ref{TDHFB}) play a role of
Kohn-Sham-like wave functions, which serve the purpose of generating
generalized density matrices ${\cal R}(t)$ only. Neither
$|\Psi\rangle$ represents a correct ground state of the system nor
$|\tilde{\Psi}\rangle$ represents that of an excited vibrational
state. However, the amplitude $\tilde{\cal R}$, which constitutes the
fundamental degree of freedom of the QRPA method, does represent a
fair approximation to the transition density matrix between both
states of the system. It then allows for calculating matrix elements
of arbitrary one-body operators between the ground state and
vibrational state, which is the primary goal of the QRPA approach.

Equation~(\ref{QRPA1}) constitutes the base for our solution of the QRPA
equations in terms of the iterative Arnoldi method. Indeed, since the
mean-field amplitude $\tilde{\cal H}$ depends linearly on the density
amplitude $\tilde{\cal R}$, Eq.~(\ref{QRPA1}) constitutes an
eigen-equation determining $\tilde{\cal R}$ and $\hbar \omega$.
However, the matrix to be diagonalized, that is the QRPA matrix, does
not have to be explicitly determined. To obtain the entire QRPA
strength function, it is enough to start from a pivot amplitude and
repeatedly act on it with the expression on the right-hand
side~\cite{[Toi10]}. In each iteration, one only has to calculate the
mean-field amplitude $\tilde{\cal H}$ corresponding to the current
density amplitude $\tilde{\cal R}$, which is an easy task.  The pivot
can be freely chosen to optimally suit the calculation.  It can for
example be random, a QRPA eigen-phonon or be constructed from an external
field.  In this work we construct the pivot from the monopole
transition operator.  This approach is fundamentally different than that used within the FAM of
Ref.~\cite{[Avo11]}, where an external field is used throughout the
calculation and Eq.~(\ref{QRPA1}) has to be iterated for all values of frequencies
$\omega$.

Since both stationary (${\cal R}^2 = {\cal R}$) and time-dependent,
(${\cal R}^2(t) = {\cal R}(t)$) density matrices are projective,
the QRPA amplitude $\tilde{\cal R}$ has vanishing matrix
elements between the quasihole and between the quasiparticle states, that is,
\begin{equation}
\varphi^{+} \tilde{\cal R} \varphi = \chi^{+} \tilde{\cal R} \chi = 0 .
\end{equation}
Therefore, $\tilde{\cal R}$ is solely defined through the antisymmetric amplitude
matrices $\tilde{Z}$ and $\tilde{Z}'^{+}$ defined as
\begin{eqnarray}
\nonumber
\tilde{Z} = - \tilde{Z}^{T} = \chi^{+} \tilde{\cal R} \varphi , \\
\tilde{Z}'^{+} = - \tilde{Z}'^{*} = \varphi^{+} \tilde{\cal R} \chi .
\end{eqnarray}
Explicitly, amplitudes $\tilde{Z}$ and $\tilde{Z}'^{+}$ read
\begin{eqnarray}
\nonumber
\tilde{Z} &=& U^{+} \tilde{\rho} V^{*} + U^{+} \tilde{\kappa} U^{*} + V^{+} \tilde{\kappa}'^{+} V^{*} - V^{+} \tilde{\rho}^{T} U^{*} ,\\
\tilde{Z}'^{+} &=& V^{T} \tilde{\rho} U + V^{T} \tilde{\kappa} V + U^{T} \tilde{\kappa}'^{+} U - U^{T} \tilde{\rho}^{T} V .
\label{ZZ1}
\end{eqnarray}
Within such a formalism, the QRPA equations (\ref{QRPA1}) can be expressed as
\begin{eqnarray}
\nonumber
-\hbar \omega \tilde{Z} &=& E \tilde{Z} + \tilde{Z} E + \tilde{W} ,\\
\hbar \omega  \tilde{Z}'^{+} &=& E \tilde{Z}'^{+} + \tilde{Z}'^{+} E + \tilde{W}'^{+} ,
\label{QRPA2}
\end{eqnarray}
where the field amplitudes $\tilde{W}$ and $\tilde{W}'^{+}$ are defined as
\begin{eqnarray}
\nonumber
\tilde{W} = - \tilde{W}^{T} = \chi^{+} \tilde{\cal H} \varphi , \\
\tilde{W}'^{+} = - \tilde{W}'^{*} = \varphi^{+} \tilde{\cal H} \chi ,
\end{eqnarray}
or explicitly,
\begin{eqnarray}
\nonumber
\tilde{W} &=& U^{+} \tilde{h} V^{*} + U^{+} \tilde{\Delta} U^{*} + V^{+} \tilde{\Delta}'^{+} V^{*} - V^{+} \tilde{h}^{T} U^{*} ,\\
\tilde{W}'^{+} &=& V^{T} \tilde{h} U + V^{T} \tilde{\Delta} V + U^{T} \tilde{\Delta}'^{+} U - U^{T} \tilde{h}^{T} V .
\end{eqnarray}
We can also invert Eq.~(\ref{ZZ1}) and obtain transition
densities $\tilde{\rho}$, $\tilde{\kappa}$, and $\tilde{\kappa}'^{+}$
expressed in terms of amplitudes $\tilde{Z}$ and $\tilde{Z}'^{+}$, that is,
\begin{eqnarray}
\nonumber
\tilde{\rho} = U \tilde{Z} V^{T} + V^* \tilde{Z}'^{+} U^{+} , \\
\nonumber
\tilde{\kappa} = U \tilde{Z} U^{T} + V^* \tilde{Z}'^{+} V^{+} , \\
\tilde{\kappa}'^{+} = V \tilde{Z} V^{T} + U^* \tilde{Z}'^{+} U^{+} .
\end{eqnarray}

Finally, we can reduce the above QRPA formalism to spherical symmetry
used in the present study. Then, the vibrating amplitude of
Eq.~(\ref{TDHFB}) has good angular-momentum quantum numbers $JM$,
that is, $|\tilde{\Psi}\rangle\equiv|\tilde{\Psi}^{JM}\rangle$ and
hence all the QRPA amplitudes pertain to the given preselected
channel $JM$, while the ground state $|\Psi\rangle$ is spherical.
As a consequence, as dictated by the angular-momentum algebra,
only specific spherical single-particle states
are coupled by the QRPA amplitudes, which can be expressed
through the Wigner-Eckart theorem and reduced matrix elements as
\begin{eqnarray}
\label{Wigner-Eckart1} \!\!\!\!
  \tilde{X}^{JM}_{\alpha jm,\alpha'j'm'}
  &\!\!=\!\!& \frac{1}{\sqrt{2j+1}} C^{jm}_{j'm'JM}
  \langle\psi_{\alpha j}||\tilde{X}^{J}||\psi_{\alpha'j'}\rangle ,
\end{eqnarray}
where $\tilde{X}$ stands for amplitudes $\tilde{\rho}$ or $\tilde{h}$,
and
\begin{eqnarray}
\label{Wigner-Eckart3a}  \!\!\!\!\!\!\!\!
  \tilde{X}^{JM}_{\alpha jm,\alpha'j'm'}
  &\!\!=\!\!& \frac{(-1)}{\sqrt{2J+1}}
C^{JM}_{jmj'm'} \langle\psi_{\alpha j}||\tilde{X}^{J}||\psi_{\alpha'j'}\rangle , \\
\label{Wigner-Eckart7a}  \!\!\!\!\!\!\!\!
  \tilde{X}'{}^{+JM}_{\alpha jm,\alpha'j'm'}
  &\!\!=\!\!& \frac{(-1)^{J-M}}{\sqrt{2J+1}}
C^{J,-M}_{jmj'm'} \langle\psi_{\alpha j}||\tilde{X}'{}^{+J}||\psi_{\alpha'j'}\rangle ,
\end{eqnarray}
where $\tilde{X}$ stands for amplitudes
$\tilde{\kappa}$, $\tilde{\Delta}$, $\tilde{Z}$, or $\tilde{W}$.
In these expressions, we have used the standard quantum numbers $\alpha jm$
of spherical single-particle states.

Spurious QRPA mode appears in the $0^+$ QRPA calculations.  In a
self-consistent full QRPA diagonalization, the spurious mode decouples
from the physical QRPA modes and appears at zero energy.  In
the Arnoldi method, this separation does not happen unless we make the
full Arnoldi diagonalization, which usually is not feasible.

To prevent the mixing of physical QRPA excitations with the spurious
$0^+$ mode, before the Arnoldi iteration we create the spurious-mode
QRPA amplitudes and its associated conjugate-state (boost-mode) QRPA
amplitudes.  The spurious $0^+$ mode amplitudes follow from the
particle number operator and have the form,
\begin{eqnarray}
  \label{eq:spurious-1}
  \tilde{P}^{00} &=&
  U^+ V^* , \quad
  \tilde{P}'^{+00} =
  V^T U.
\end{eqnarray}
The $0^+$ boost mode is generated by making an additional HFB
calculation whose chemical potentials $\lambda_\tau$ and average
particle numbers are slightly shifted from the ground state values,
producing a perturbed state $\ket{{\rm HFB}_2}$.  The boost-mode
amplitudes are calculated by using Thouless theorem as,
\begin{eqnarray}
  \label{eq:boost-1}
  \tilde{R}^{00}_{\alpha jm, \alpha' j'm'} &=& \frac{
  \bra{{\rm HFB}_2} \ad{\alpha jm} \ad{\alpha'j'm'} \ket{{\rm HFB}} }
{ \ovlap{{\rm HFB}_2}{{\rm HFB}} }
\nonumber\\
  &=& \bigl( {\tilde V} {\tilde U}^{-1} \bigr)^{}_{\alpha jm, \alpha' j'm'},
\\
  \label{eq:boost-1a}
  \tilde{R}^{'+00}_{\alpha jm, \alpha' j'm'} &=& \frac{
  \bra{{\rm HFB}} a_{\alpha' j'm'} a_{\alpha jm}   \ket{{\rm HFB}_2} }
{ \ovlap{{\rm HFB}}{{\rm HFB}_2} }
\nonumber\\
  &=& \bigl( {\tilde V} {\tilde U}^{-1} \bigr)^*_{\alpha jm, \alpha' j'm'},
\end{eqnarray}
where we used the standard transformation matrices from one
quasiparticle basis to another~\cite{[RS80]},
\begin{eqnarray}
  \label{eq:boost-2}
  {\tilde V} &=& U^T V_2 + V^T U_2, \\
  {\tilde U} &=& U^+ U_2 + V^+ V_2.
\end{eqnarray}

Gram-Schmidt orthogonalization is used to keep during the Arnoldi
iteration the Krylov-space basis vectors orthogonal to the spurious
and boost modes, that is, each Krylov-space basis vector is orthogonalized
against ${\tilde P}$ and ${\tilde R}$. The orthogonalization
procedure is described in detail in Ref.~\cite{[Toi10]}. For the
semi-magic nuclei considered here, we only vary the particle number of
the nucleon species that has non-vanishing pairing correlations.

\section{Separable Pairing Interaction}
\label{separable}

The separable finite-range pairing interaction for neutrons ($\tau=n$)
and protons ($\tau=p$) that we use in this
study is defined as~\cite{[Tia09]}
\begin{eqnarray}
 \nonumber
 \hat{V}_\tau (\bm{r}_1 s_1, \bm{r}_2 s_2; \bm{r}'_1 s'_1, \bm{r}'_2 s'_2)
\hspace*{-3cm} &&\nonumber \\
&=& - G_\tau \delta (\bm{R}-\bm{R}') P(r) P(r')
\frac{1}{2} (1-\hat{P}_{\sigma}) ,
\label{sep-int}
\end{eqnarray}
where $\bm{R}=(\bm{r}_1+\bm{r}_2)/2$ denotes the centre of mass
coordinate, $\bm{r}=\bm{r}_1-\bm{r}_2$ is the relative coordinate,
$r=|\bm{r}|$, $\hat{P}_{\sigma}$ is the standard
spin-exchange operator, and function $P(r)$ is a sum of $m$ Gaussian terms,
\begin{equation}
P(r)=\frac{1}{m} \sum_{i=1}^{m} \frac{1}{(4\pi a_i^2)^{3/2}}
\mbox{e}^{-\frac{r^2}{4a_i^2}} .
\label{Gauss}
\end{equation}
Coupling constants $G_\tau$ define the pairing strengths for neutrons and protons.

For such a pairing interaction, the pairing energy acquires a fully separable form,
which in spherical symmetry reads
\begin{eqnarray}
 \nonumber
 E_{\text{pair}}^{\text{sep}} &=&
 - \frac{1}{2} \sum_{NJ\tau} G_\tau \Bigl( \sum_{\mu \nu} V_{\mu \nu}^{NJ}
\langle\psi_{\mu}||\kappa'^{+J}_\tau||\psi_{\nu}\rangle \Bigr) \\
&&\hspace*{1.5cm}\times\Bigl( \sum_{\mu' \nu'} V_{\mu' \nu'}^{NJ}
\langle\psi_{\mu'}||\kappa^{J}_\tau||\psi_{\nu'} \rangle \Bigr) ,
\label{Pair-En}
\end{eqnarray}
and depends on the reduced matrix elements of the pairing densities
$\kappa_\tau$ and $\kappa'^{+}_\tau$ between the single-particle wave
functions $\psi_\mu(\bm{r})$ for $\mu$ denoting the set of spherical
harmonic-oscillator (HO) quantum numbers $n_\mu l_\mu j_\mu$. The interaction matrix elements
$V_{\mu \nu}^{NJ}$ are defined as
\begin{eqnarray}
 \nonumber
 V_{\mu \nu}^{NJ} &=& \sqrt{(4\pi)(2J+1)(2j_{\mu}+1)(2j_{\nu}+1)}
 \begin{Bmatrix} l_{\mu} & l_{\nu} & J \\ \frac{1}{2} & \frac{1}{2} &
 0 \\ j_{\mu} & j_{\nu} & J \end{Bmatrix}
 \\ \nonumber
&\times&  M^{NJn0}_{n_{\mu}l_{\mu}n_{\nu}l_{\nu}} \frac{2^{1/4}}{b^{3/2}}
\sqrt{\frac{\pi^{1/2}(2n+1)!}{2(2^nn!)^2}}
 \frac{1}{m} \sum_{i=1}^{m} \frac{1}{(4\pi a_i^2)^{3/2}}  \\
&\times& \Bigl(
\frac{2a_i^2b^2}{1+a_i^2b^2} \Bigr)^{3/2}
\Bigl( \frac{1-a_i^2b^2}{1+a_i^2b^2} \Bigr)^{n} ,
\label{V-isoscal}
\end{eqnarray}
where $2n=2n_{\mu}+l_{\mu}+2n_{\nu}+l_{\nu}-2N-J$,
$M^{N\lambda n0}_{n_{\mu}l_{\mu}n_{\nu}l_{\nu}}$ are the standard
Talmi-Moshinski coefficients~\cite{[Bar66]}, and $b=\sqrt{m\omega/\hbar}$ denotes the
HO constant.

\section{Nuclear Incompressibility}
\label{sec4}

The isoscalar incompressibility of infinite nuclear matter is defined by
the well-known formula~\cite{[Bla80]}
\begin{equation}
K_{\infty} = 9\rho^2 \frac{\mbox{d}^2}{\mbox{d}\rho^2}
\Bigl( \frac{E}{A} \Bigr)_{\rho=\rho_{nm}},
\label{NM_eq}
\end{equation}
where $\rho_{nm}$ is the saturation density of nuclear matter. Of
course, $K_{\infty}$ cannot be directly measured; however, by using
Eq.~(\ref{NM_eq}) it can be calculated from theoretical equation of
state $E(\rho)$ or it can be indirectly estimated from measurements
of monopole excitations of finite nuclei.

The incompressibility of finite nucleus, $K_A$, is defined by its scaling-model
relation~\cite{[Str82]}  to
the centroid of the giant monopole resonance (GMR), $E_{\text{GMR}}$, as
\begin{equation}
E_{\text{GMR}} = \sqrt{\frac{\hbar^2K_A}{m\langle r^2\rangle}}
\label{KA}
\end{equation}
where $\langle r^2\rangle$ is the average square radius of the nucleus.
Eq.~(\ref{KA}) is derived under the assumption that most of the monopole
strength is concentrated within one dominant peak, see Ref.~\cite{[Bla80]}.
However, often the monopole giant resonances consist of more than
one dominant peak. The reliability of the scaling-model was
also challenged in, e.g., Ref.\cite{[Nis85]}. Therefore, we want to
emphasize that extracting the incompressibility $K_A$ from the GMR
centroid in Eq.~(\ref{KA}) is only approximative and model-dependent.
For this reason, we pay attention to analyze not only the nuclear
incompressibilities, but also directly the GMR centroids.

The centroid of the GMR can be extracted from its strength function as
the ratio of the first and zero moments, that is,
\begin{equation}
E_{\text{GMR}} = \frac{m_1}{m_0}.
\label{centroid}
\end{equation}
There exist several alternative ways to extract $E_{\text{GMR}}$ through
different moments of the strength function, such as
$E_{\text{GMR}}=\sqrt{m_1/m_{-1}}$ or
$E_{\text{GMR}}=\sqrt{m_3/m_{1}}$. However, they are more sensitive
to details of the strength function and thus less appropriate
for studies of the incompressibility.

In analogy to the Weizs\"{a}cker formula for the nuclear masses,
one can introduce~\cite{[Bla80]} a similar relation for nuclear incompressibilities,
\begin{eqnarray}
 K_A &=& K_{V} + K_{S}A^{-1/3} + (K_{\tau}+K_{S,\tau}A^{-1/3})\frac{(N-Z)^2}{A^2}
\nonumber \\
 &&+ K_{C} \frac{Z^2}{A^{4/3}} \ .
 \label{liquid-drop}
\end{eqnarray}
Similarly as in the liquid-drop (LD) model, we refer to $K_{V}$, $K_S$, $K_{\tau}$,
$K_{S,\tau}$, and $K_C$ as the volume, surface, symmetry, surface-symmetry,
and Coulomb incompressibility parameters, respectively. By adjusting
these parameters to the incompressibilities $K_A$, calculated in finite nuclei
from Eqs.~(\ref{KA}) and (\ref{centroid}),
we can obtain an estimate of the infinite-matter incompressibility as
$K_{\infty}\simeq K_{V}$.

\section{Results}
\label{sec5}

In our study we performed a set of calculations for semi-magic
nuclei starting from $Z=8$ or $N=8$ and ending with $Z=82$ or
$N=126$. The ground states properties were calculated within the
HFB method by using the code HOSPHE~\cite{[Car10d]},
whereas the monopole strength functions
were obtained by implementing in the same code the QRPA method
within the Arnoldi iterative method~\cite{[Toi10]}.

We decided to use two different Skyrme functionals --
SLy4~\cite{[Cha98]} and UNEDF0~\cite{[Kor10b]}. Both of them were
tuned (among other observables) to reproduce the main properties of
the infinite nuclear matter. In particular, they correspond
to the same value of nuclear incompressibility (\ref{NM_eq})
of $K_{\infty}=230$\,MeV and differ in their values of the effective
mass of $m^*/m=0.70$ and 1.11 for SLy4 and UNEDF0, respectively.

The present study is focused on comparing incompressibilities
obtained with two different pairing interactions, namely, the
standard zero-range force,
$V_\tau(\bm{r},\bm{r}')=-V_{0\tau}
\delta(\bm{r}-\bm{r}')$, and separable force presented in
Sec.~\ref{separable}. To make the comparison meaningful, we adjusted
the strength parameters, $G_\tau$ and $V_\tau$, so as to obtain for
both forces very similar neutron (proton) pairing gaps in $Z=50$
isotopes ($N=50$ isotones). The resulting gaps roughly correspond to
the experimental odd-even mass staggering along the $Z=50$ and $N=50$
chains of nuclei. Theoretical pairing gaps, $\Delta_n$ and $\Delta_p$, were
determined as in Ref.~\cite{[Dob96]}, namely,
\begin{equation}
\Delta_\tau = \frac{\mbox{Tr}' (\rho_{\tau} \Delta_{\tau})}{\mbox{Tr} \rho_{\tau}}
\label{gap-th}
\end{equation}
where $\mbox{Tr} A= \sum_k A_{kk}$ and $\mbox{Tr}' A = \sum_{k>0}
A_{k\bar{k}}$. For the separable pairing, in Eq.~(\ref{Gauss}) we
used only one Gaussian term with $a_1=0.66$\,fm.

\begin{figure}[ht]
 \begin{center}
   \includegraphics[width=0.9\columnwidth,angle=0]{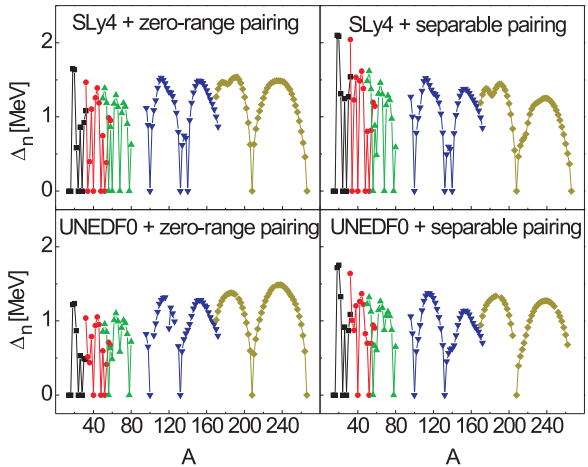}
  \end{center}
\caption[F]{(Color online) Neutron pairing gaps in the $Z=8$, 20, 28, 50, and 82
isotopes (see the legend shown in Fig.~\protect\ref{fig3}).
Upper and lower panels show
results obtained for the SLy4 and UNEDF0 functionals,
respectively. Left and right panels show results obtained for the
zero-range and separable pairing, respectively.}
\label{fig1}
\end{figure}

\begin{figure}[ht]
 \begin{center}
   \includegraphics[width=0.9\columnwidth,angle=0]{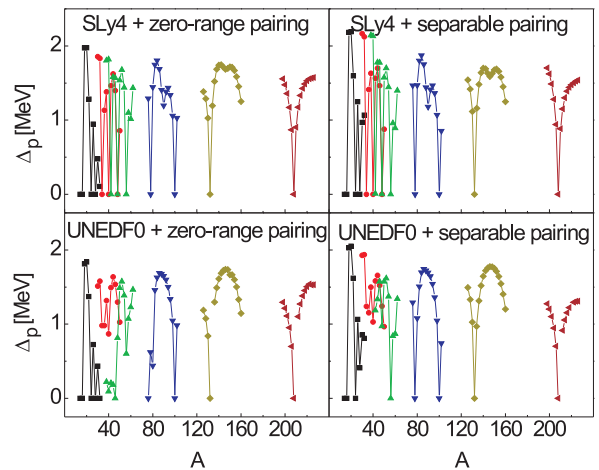}
  \end{center}
\caption[F]{(Color online) Same as in Fig.~\protect\ref{fig1} but for the proton gaps
in the $N=8$, 20, 28, 50, 82, and 126 isotones (see the legend shown
in Fig.~\protect\ref{fig4}).}
\label{fig2}
\end{figure}

In this way, in the calculations we used
the separable-force strength parameters of $G_n=631$
and 473\,MeV\,fm$^3$ ($G_p=647$ and 521\,MeV\,fm$^3$) for
the SLy4 and UNEDF0 functionals, respectively, and similarly,
for the zero-range force: $V_n=195$ and 126\,MeV\,fm$^3$
($V_p=221$ and 157\,MeV\,fm$^3$). All calculated
neutron and proton pairing gaps are shown in Figs.~\ref{fig1}
and~\ref{fig2}, respectively. One can see that the results obtained
for both pairing forces are fairly similar. The HFB iterations were
carried out using a linear mixing of densities from the current and
previous iteration defined by a constant mixing parameter
\cite{[Car10d]}. With this recipe, for some of the nuclei, the HFB
iterations did not end in converged solutions. Such cases were
excluded from the analysis of pairing properties and the subsequent
QRPA calculations.

We note here that no energy cut-off is needed for calculations using
the separable force, and thus in our calculations the entire
HO basis up to $N_0=20$ shells was used, see
Appendix~\ref{app1}. On the other hand, for the zero-range force we
used the cut-off energy of 60\,MeV applied within the two-basis
method \cite{[Gal94],[Sch12]}.

\begin{figure}[ht]
 \begin{center}
   \includegraphics[width=0.9\columnwidth,angle=0]{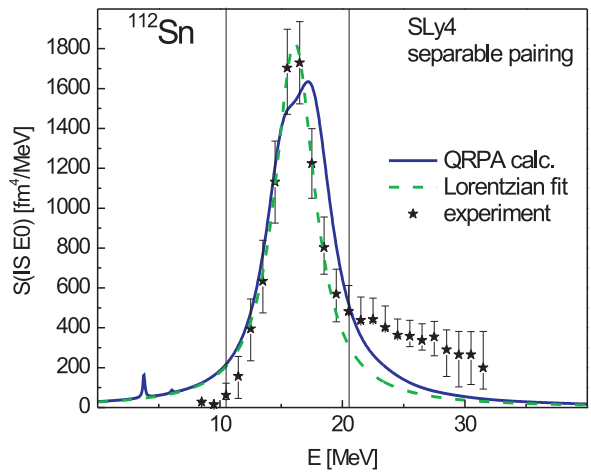}
  \end{center}
  \caption[F]{(Color online) The QRPA monopole strength function in $^{112}$Sn
(solid line) compared to raw experimental data~\cite{[Li10]} and
Lorentzian fit to data (dashed line) performed in the region of
energies of 10.5--20.5\,MeV~\cite{[Li10]}. }
\label{fig12}
 \end{figure}

In Fig.~\ref{fig12} we compare our QRPA results with raw experimental
data obtained in Ref.~\cite{[Li10]}. In this work, a Lorentzian fit to data was performed in
the region of energies of 10.5--20.5\,MeV, and the experimental
values of ${m_1}/{m_0}$ were determined from the corresponding
fitted curve (its moments were calculated for energies from zero
to infinity). In determining our theoretical values of ${m_1}/{m_0}$,
we also perform the integration in the entire energy domain. We have
checked that the integration of theoretical curves in the fixed
region of 10.5--20.5\,MeV does not bring meaningful results, because,
in the wide region of masses studied here, the GMR peaks move too much, and extend
beyond the above narrow range of energies. Our QRPA strength
functions were obtained from the discrete Arnoldi strength
distributions by using the smoothing methods explained in
Ref.~\cite{[Toi10]}. We also note that in our QRPA calculations, the
high-energy shoulder of the strength function is not obtained, cf.\
discussion in Ref.~\cite{[Li10]}.

\begin{figure}[ht]
 \begin{center}
   \includegraphics[width=0.9\columnwidth,angle=0]{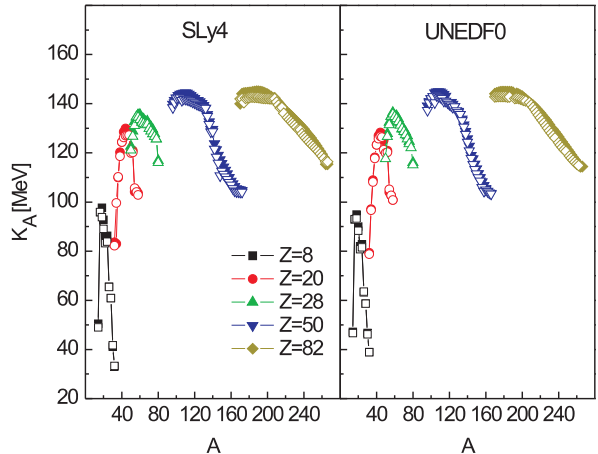}
  \end{center}
  \caption[F]{(Color online) Incompressibility $K_A$ calculated for the
isotopic chains of semimagic nuclei with $Z=8$, 20, 28, 50, and 82.
Left and right panels show results obtained for
the SLy4 and UNEDF0 functionals, respectively. Full (empty) symbols
correspond to the zero-range (separable) pairing force.}
\label{fig3}
\end{figure}

\begin{figure}[ht]
 \begin{center}
   \includegraphics[width=0.9\columnwidth,angle=0]{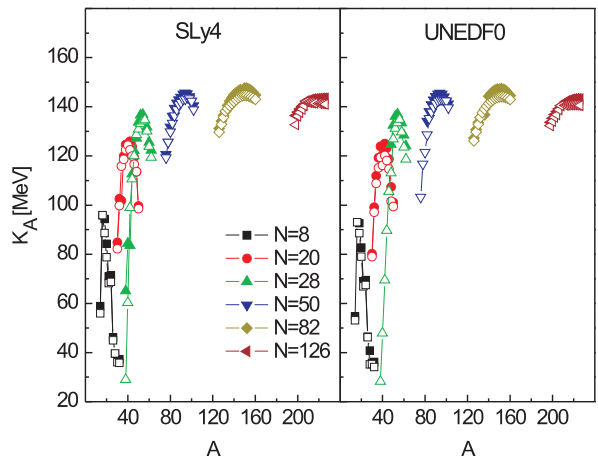}
  \end{center}
 \caption[F]{(Color online) Same as in Fig.~\protect\ref{fig3}, but for the isotonic chains with
$N=8$, 20, 28, 50, 82, and 126.}
\label{fig4}
 \end{figure}

Figs.~\ref{fig3} and \ref{fig4} present the overview of all obtained
finite-nucleus incompressibilities $K_A$,
Eqs.~(\ref{KA}) and (\ref{centroid}), calculated along the isotopic and
isotonic chains, respectively. One can see that for both Skyrme
functionals, SLy4 and UNEDF0, values corresponding to the
zero-range (full symbols) and separable (open symbols) pairing
forces are very similar.

\begin{figure}[ht]
 \begin{center}
   \includegraphics[width=0.9\columnwidth,angle=0]{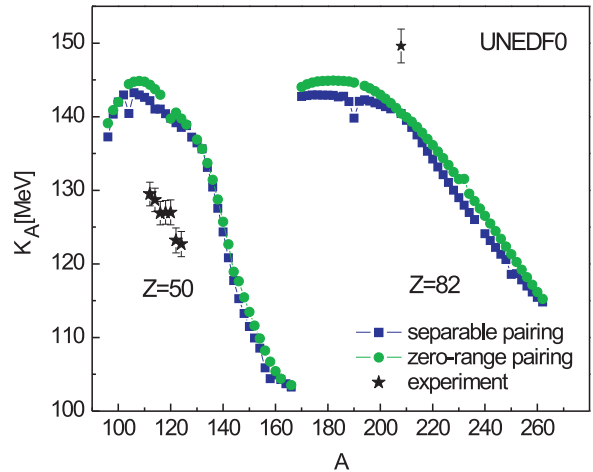}
  \end{center}
  \caption[F]{(Color online) Incompressibility $K_A$ calculated for
chains of the $Z=50$ and 82 isotopes. Results obtained by using the separable
(squares) and zero-range (circles) pairing with the UNEDF0 functional are compared to the
available experimental data~\protect\cite{[You99],[Li07],[Li10]}.}
\label{fig5}
 \end{figure}

\begin{figure}[ht]
 \begin{center}
   \includegraphics[width=0.9\columnwidth,angle=0]{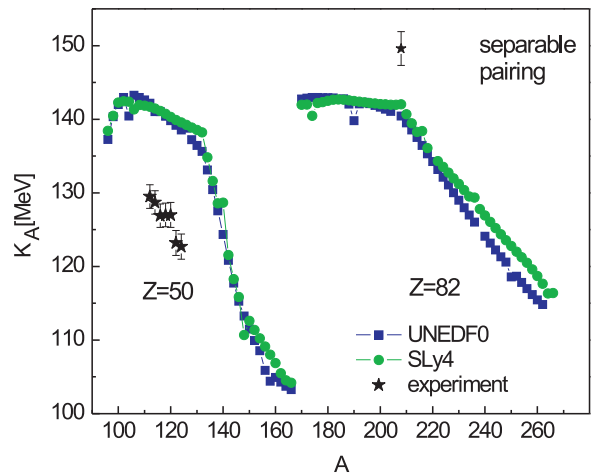}
  \end{center}
  \caption[F]{(Color online) Same as in Fig.~\protect\ref{fig5}, but for the UNEDF0
(squares) and SLy4 (circles) functionals and separable pairing force.}
\label{fig8}
 \end{figure}

To see effects of the pairing interaction in more detail, we
focus on the results obtained for chains of tin and lead isotopes. In
Figs.~\ref{fig5} and \ref{fig8} we compare theoretical results with
the experimental data for $^{208}$Pb and $^{112-124}$Sn, taken from Refs.~\cite{[You99],[Li07],[Li10]}.
A comparison of the two types of pairing interactions, and two different
Skyrme functionals, leads to the conclusion that the calculated
incompressibilities $K_A$ depend on
the interactions in the particle-particle channel as well as the
particle-hole channel of the two Skyrme functionals used in our study -
SLy4 and UNEDF0 - only weakly.
Of course, we can expect that using Skyrme parametrizations
tuned to higher (lower) values of $K_{\infty}$ may lead to
uniformly higher (lower) values of $K_A$.

\begin{figure}[ht]
 \begin{center}
   \includegraphics[width=0.9\columnwidth,angle=0]{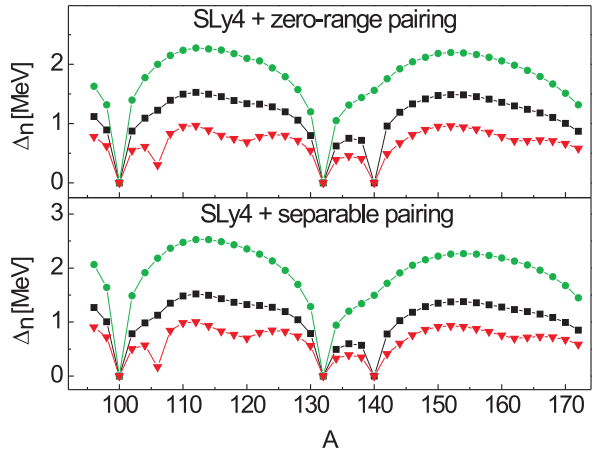}
  \end{center}
  \caption[F]{(Color online) Neutron pairing gaps calculated in tin
isotopes for low (triangles), central (squares), and high (circles)
values of pairing strength parameters given in captions of
Figs.~\protect\ref{fig6} and~\protect\ref{fig9}.}
\label{fig13}
 \end{figure}

\begin{figure}[ht]
 \begin{center}
   \includegraphics[width=0.9\columnwidth,angle=0]{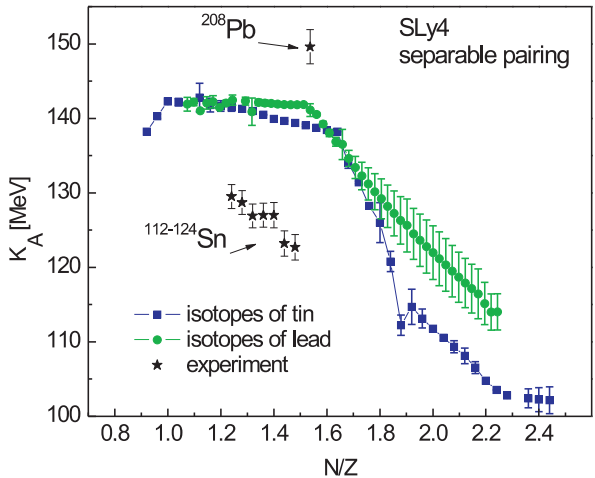}
  \end{center}
  \caption[F]{(Color online)
Incompressibility $K_A$ calculated for the SLy4 functional and
separable pairing force in tin (squares) and lead (circles) isotopes
compared to the available experimental data. Theoretical results are
plotted together with uncertainties pertaining to variations of the
neutron strength parameter in the range of
$G_n=631\pm150$\,MeV\,fm$^3$.}
\label{fig6}
 \end{figure}

\begin{figure}[ht]
 \begin{center}
   \includegraphics[width=0.9\columnwidth,angle=0]{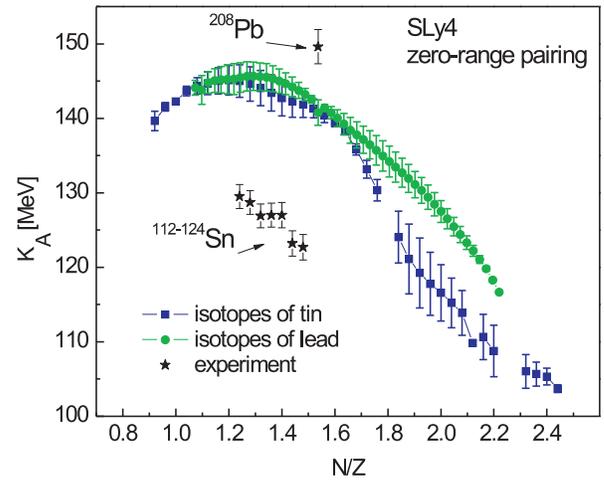}
  \end{center}
  \caption[F]{(Color online) Same as in Fig.~\protect\ref{fig6},
but for the zero-range pairing force and uncertainties pertaining to variations of the
neutron strength parameter in the range of $V_n=195\pm30$\,MeV\,fm$^3$.}
\label{fig9}
 \end{figure}

To check a weak dependence of $K_A$ on the intensity of pairing
correlations, we have repeated the calculations by using values of
neutron pairing strengths varied in a wide range,
$G_n=631\pm150$\,MeV\,fm$^3$ and $V_n=195\pm30$\,MeV\,fm$^3$. Such
variations induce very large changes of neutron pairing gaps, shown
in Fig.~\ref{fig13}; the ones that are certainly beyond any
reasonable range of uncertainties related to adjustments of
pairing strengths to data. In Figs.~\ref{fig6} and \ref{fig9}, we
show the influence of the varied pairing strengths on the calculated
incompressibilities $K_A$. We see clearly that even such large
variations cannot induce changes compatible with discrepancies with
experimental data.

To illustrate the effect of isospin asymmetry, in Figs.~\ref{fig6}
and \ref{fig9} we plotted the results as functions of $N/Z$, whereby
$^{124}$Sn and $^{208}$Pb are located at almost the same point of the
abscissa. These figures clearly show that the discrepancies with data
are probably not related to the isospin dependence of $K_A$. Indeed,
for both types of pairing, in the region of $1.0 < N/Z < 1.6$, the
results obtained for tin and lead isotopes roughly follow each other.

\begin{figure}[ht]
 \begin{center}
   \includegraphics[width=0.9\columnwidth,angle=0]{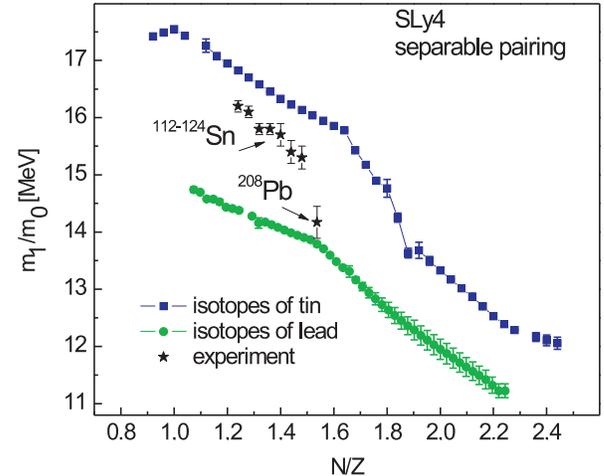}
  \end{center}
  \caption[F]{(Color online) Same as in Fig.~\protect\ref{fig6},
but for the centroids $m_1/m_0$.}
\label{fig7}
 \end{figure}

\begin{figure}[ht]
 \begin{center}
   \includegraphics[width=0.9\columnwidth,angle=0]{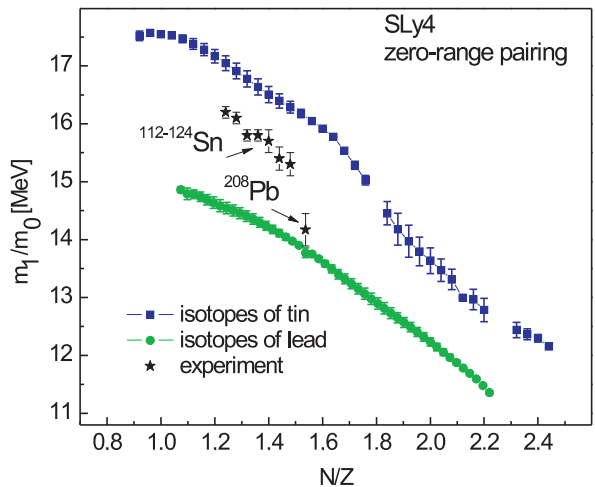}
  \end{center}
  \caption[F]{(Color online) Same as in Fig.~\protect\ref{fig7},
but for the zero-range pairing force.}
\label{fig10}
 \end{figure}

Finally, to illustrate the fact that nuclear radii are fairly robust
and cannot significantly influence the values of $K_A$, determined from
Eqs.~(\ref{KA}) and (\ref{centroid}), we show values of $m_1/m_0$
alone in Figs.~\ref{fig7} and \ref{fig10}.  We see that for both
types of pairing, in tin and lead the calculated values of $m_1/m_0$
overestimate and underestimate the measured ones by 0.6--0.8 and
0.4\,MeV, respectively. Exactly the same pattern was obtained within
the relativistic nuclear energy density functionals studied in
Ref.~\cite{[Nik08a]}, where the corresponding discrepancies were
equal to 0.8--1.0 and 0.2\,MeV. We also note that this comparison
directly relates calculations to data, without using the intermediate
and model-dependent definition of $K_A$.

To conclude our analysis, we have performed adjustments of the
LD formula (\ref{liquid-drop}) to our microscopically
calculated values of $K_A$.
Since in the LD formula all parameters appear linearly, we could use
the standard linear-regression method, which gave us the values of
parameters that minimize $\chi^2$ along with standard estimates of
statistical errors.

The obtained parameters are collected in
Table~\ref{tab1}. We see that the LD formula is able to
provide an excellent description of the QRPA results, with
average deviations of the order of 5\,MeV, that is,
about 3\% of the typical value of $K_A$. Similarly the values of the
volume incompressibility $K_V$ are determined to about 2\% of
precision. The least precisely determined LD parameter is the
surface-symmetry incompressibility $K_{S,\tau}$, estimated up to 25\%
of precision. We also note that, within the fit precision, the volume
parameter $K_V$ averaged over both functionals and both pairing
forces equals to
254$\pm$5\,MeV, which is significantly higher than the corresponding
infinite-matter incompressibility of $K_{\infty}$=230\,MeV.
We would like to point out that the errors given in
Table~\ref{tab1} are the statistical errors of the adjusted parameters
and do not take into account
possible systematic errors caused by using the model-dependent
Eq.~(\ref{KA}). Nevertheless, the results of the fit can be used
as a useful parameterization of the microscopic calculations.
\newlength{\di}
\setlength{\di}{4mm}
\begin{table}
\caption[T]{\label{tab1} Parameters (in MeV) of the LD
formula (\protect\ref{liquid-drop}) with standard errors, obtained by
a fit to the values of $K_A$ calculated in $M$ semi-magic nuclei
across the mass chart. The parameter $\chi$ was determined as the
square root of the sum of fit residuals squared divided by the number
of fit degrees of freedom ($M-5$ in our case).}
\begin{center}\begin{tabular}{l@{\hspace{\di}}|r@{}c@{}l@{\hspace{\di}}r@{}c@{}l@{\hspace{\di}}r@{}c@{}l@{\hspace{\di}}r@{}c@{}l}
\hline
\hline
&\multicolumn{6}{c}{SLy4}&\multicolumn{6}{c}{UNEDF0}  \\
%\cline{2-13}
&\multicolumn{3}{c}{separable} & \multicolumn{3}{c}{zero-range} & \multicolumn{3}{c}{separable} & \multicolumn{3}{c}{zero-range}  \\
\hline
$K_{V}$     &    252&$\pm$&5   &    258&$\pm$&5   &    249&$\pm$&5   &    257&$\pm$&4    \\
$K_{S}$     & $-$391&$\pm$&14  & $-$406&$\pm$&13  & $-$397&$\pm$&14  & $-$412&$\pm$&13   \\
$K_{\tau}$  & $-$460&$\pm$&30  & $-$500&$\pm$&30  & $-$510&$\pm$&30  & $-$550&$\pm$&30   \\
$K_{S,\tau}$&    410&$\pm$&110 &    560&$\pm$&100 &    570&$\pm$&120 &    740&$\pm$&100  \\
$K_{C}$     & $-$5.2&$\pm$&0.4 & $-$5.4&$\pm$&0.4 & $-$4.5&$\pm$&0.4 & $-$5.1&$\pm$&0.4  \\
\hline
$M$           & \multicolumn{3}{c}{210}             & \multicolumn{3}{c}{211}           & \multicolumn{3}{c}{204}            & \multicolumn{3}{c}{195}             \\
$\chi$        & \multicolumn{3}{c}{5.0}             & \multicolumn{3}{c}{4.7}           & \multicolumn{3}{c}{5.3}            & \multicolumn{3}{c}{4.4}             \\
\hline
\hline
\end{tabular}\end{center}
\end{table}

\section{Conclusions}
\label{sec6}

In this work we have presented the first application of the
separable, finite-range pairing interaction of the Gaussian form
together with the non-relativistic functional of the Skyrme type. This
interaction was used to determine both the ground-state
Hartree-Fock-Bogolyubov solutions and
Quasiparticle-Random-Phase-Approximation monopole strength functions
in semi-magic nuclei. Results were systematically compared with
those pertaining to the standard zero-range pairing interaction.

From the monopole strength functions, we extracted the finite-nucleus
incompressibilities and compared them to experimental data. It turned
out that neither zero-range nor separable pairing effects were able
to describe the low values of incompressibilities measured in tin,
relative to the high value measured in $^{208}$Pb. By changing the
infinite-matter incompressibility, one can certainly describe either
the tin or lead values; however, the high difference thereof remains
unexplained.

The lack of agreement with experimental data is evident
also in the case of the GMR centroids. This is even more important
for the conclusions of our work, since the analysis of the centroids
is not affected by the model-dependent extraction of incompressibilities
by way of Eq.~(\ref{KA}).

We have also performed adjustments of the LD formula to
microscopically calculated incompressibilities, and we found that (i)
such a formula is able to describe microscopic results very well,
and (ii) the volume LD term is significantly higher than the
infinite-matter incompressibility determined for a given functional.

\section{acknowledgements}
We thank Umesh Garg for discussions regarding the experimental data.
This work was supported in part by the Academy of Finland and University
of Jyv\"{a}skyl\"{a} within the FIDIPRO programme.

\appendix

\section{Numerical tests}
\label{app1}

\begin{figure}[ht]
 \begin{center}
   \includegraphics[width=0.9\columnwidth,angle=0]{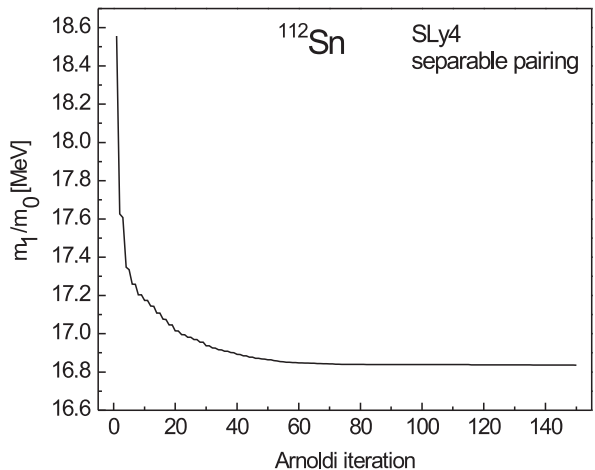}
  \end{center}
  \caption[F]{Convergence of the ratio of first and zero moments ${m_1}/{m_0}$
calculated in $^{112}$Sn as a function of the number of Arnoldi iterations.}
\label{fig11}
 \end{figure}
Fig.~\ref{fig11} illustrates the reliability of the
Arnoldi method in determining the key factors of our analysis,
namely, the ratios of moments of the monopole strength functions. To obtain a
perfectly stable result, only about 70 Arnoldi iterations suffice. In
this way, the QRPA result is achieved within the CPU time that is of
the same order as that needed to obtain a converged HFB ground state.
Note that the Arnoldi iteration conserves all odd moments, so during
the iteration, the moment ${m_1}$ does not change; thus the convergence of
${m_1}/{m_0}$ simply illustrates the convergence of ${m_0}$ alone.

\begin{figure}[ht]
\begin{center}
\includegraphics[width=0.9\columnwidth,angle=0]{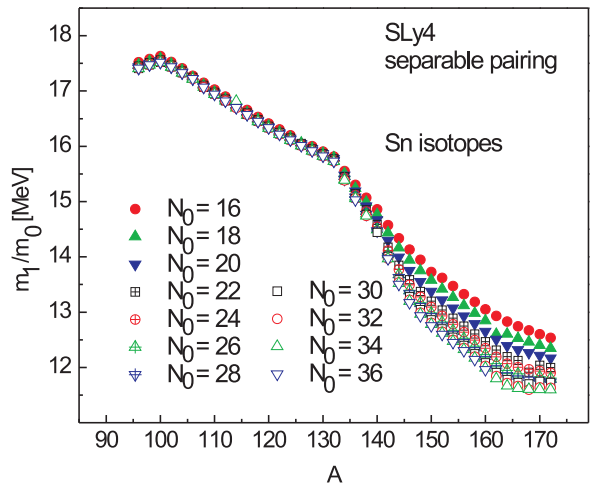}
\end{center}
\caption[F]{(Color online) Dependence of the ratio of first and zero
moments ${m_1}/{m_0}$ on the number of HO shells $N_0$,
calculated in tin isotopes.}
\label{fig14}
\end{figure}
The HO basis used in our calculations is characterized by two
numerical parameters: frequency $\hbar\omega$ and number of shells
included in the basis $N_0$. With varying particle numbers $A$, we
use the standard prescription of
\begin{equation}\label{eq:123}
\hbar\omega=1.2\times41\,\mbox{MeV}\times A^{-1/3},
\end{equation}
established for the
ground-state calculations~\cite{[Dob97d]}. Within this prescription,
in Fig.~\ref{fig14} we study dependence of the QRPA moments
${m_1}/{m_0}$ on the number of HO shells $N_0$. One can see that
in well-bound tin isotopes with $A\leq132$, one obtains
perfectly-well converged results. As is well known, in weakly-bound
isotopes, owing to the effects of coupling to the continuum, the
convergence properties gradually deteriorate and the HO-basis calculations
become less reliable.

\begin{figure}[ht]
\begin{center}
\includegraphics[width=0.9\columnwidth,angle=0]{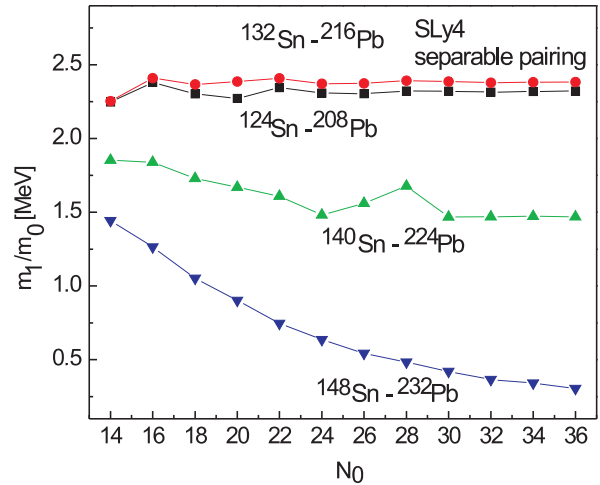}
\end{center}
\caption[F]{(Color online) Dependence of differences of ${m_1}/{m_0}$, calculated
for pairs of tin and lead isotopes, on the number of
HO shells $N_0$.}
\label{fig16}
\end{figure}
Nevertheless, as is often the case for restricted-space calculations,
results pertaining to relative observables are much less basis-dependent.
This is illustrated in Fig.~\ref{fig16}, where we show differences of
ratios of the QRPA moments ${m_1}/{m_0}$, calculated for pairs of tin
and lead isotopes. We start form the pair of well-bound isotopes,
$^{124}$Sn and $^{208}$Pb, where experimental data are known, but we
also show pairs with 8, 16, and 24 more neutrons. We see again that
results for well-bound isotopes are perfectly-well converged. However,
even for very exotic weakly-bound nuclei, the HO basis provides
reasonably reliable results.

\begin{figure}[ht]
\begin{center}
\includegraphics[width=0.9\columnwidth,angle=0]{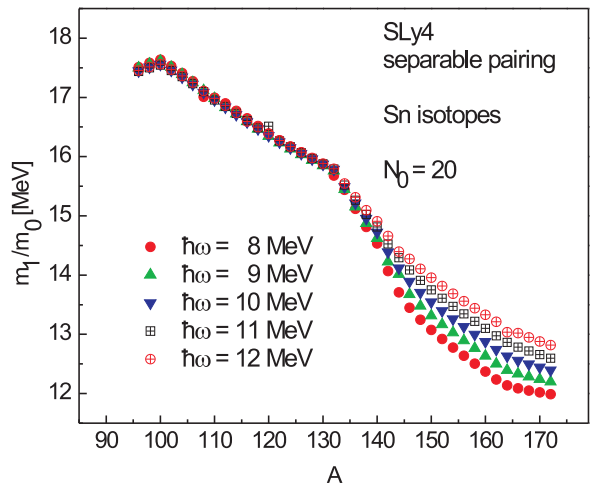}
\end{center}
\caption[F]{(Color online) Dependence of the ratio of first and zero
moments ${m_1}/{m_0}$ on the HO frequency $\hbar\omega$,
calculated in tin isotopes.}
\label{fig15}
\end{figure}
Finally, in Fig.~\ref{fig15} we show dependence of results on the
HO frequency $\hbar\omega$, determined for $N_0=20$ HO shells.
Note that the range of frequencies shown in the plot is much wider
than those corresponding to prescription (\ref{eq:123}), which gives
$\hbar\omega=10.60$ and 8.88\,MeV for $^{100}$Sn and $^{170}$Sn,
respectively. Nevertheless, no significant $\hbar\omega$-dependence
is obtained for the $A\leq132$ isotopes, whereas for weakly bound
ones the estimated uncertainty does not exceed 1\,MeV.

As an additional check, for the tin isotope $^{112}$Sn we performed
the standard QRPA calculation by using the same $N_{0} = 20$
configuration space as that used for our Arnoldi-method
calculations. We found the spurious 0$^+$ peak at a very small
energy of $5.2 \times 10^{-6}$\,MeV, which guarantees a proper
separation of the spurious mode from the physical spectrum.

%\bibliography{C:/Actual/LaTeX/Temp/jd,C:/Actual/LaTeX/Latex.all/jacwit29}
\bibliographystyle{unsrt}

\end{document}